\documentclass[12pt,a4paper]{article}
\title{Discrete Q- and P-symbols for spin $s$}
\author{Jean-Pierre Amiet and Stefan Weigert \\
Institut de Physique, Universit\'e de Neuch\^atel\\
Rue A.-L. Breguet 1, CH-2000 Neuch\^atel, Switzerland\\
\tt stefan.weigert@iph.unine.ch}
\date{June 1999}
\newcommand\be{\begin{equation}}
\newcommand\ee{\end{equation}}
\newcommand\bea{\begin{eqnarray}}
\newcommand\eea{\end{eqnarray}}
\newcommand\ket[1]{|#1\rangle}
\newcommand\bra[1]{\langle #1|}

\begin{document}
\maketitle
\begin{abstract}
Non-orthogonal bases of projectors on coherent states are introduced to expand hermitean operators acting on the Hilbert space of a spin $s$. It is shown that the expectation values of a hermitean operator $\widehat A$ in a family of $(2s+1)^2$ spin-coherent states determine the operator unambiguously. In other words, knowing the 
Q-symbol of $\widehat A$ at $(2s+1)^2$ points on the unit sphere is already sufficient in order to recover the operator.  This provides a straightforward method to reconstruct the mixed state of a spin since its density matrix is explicitly parametrized in terms of expectation values. Furthermore, a discrete P-symbol emerges naturally which is related to a basis {\em dual} to the original one.
\end{abstract}
\section{Introduction}
Phase-space formulations of quantum systems have a long history which starts with 
Wigner's introduction of a quasi-probability distribution to represent the state of a quantum particle \cite{wigner32}. Moyal has shown that not only states but also operators can be mapped onto functions on the phase space, or {\em symbols} \cite{moyal49}. In such a representation a `twisted product' of symbols keeps track of a possible non-commutativity of the underlying operators. The inverse procedure, mapping phase-space functions into operators, can be considered as quantization according to Weyl \cite{weyl28}.

A unifying approach to symbols of operators is formulated conveniently in terms of coherent states $\ket{\alpha}$ for a particle defined as  
\be
a \ket{\alpha} = \alpha \ket{\alpha} \, , \qquad \alpha \in {\sf C}\, ;
\label{CS}
\ee
here $a$ and its adjoint $a^+$, satisfying the commutation relation $[ a , a^+ ] = 1$, are annihilation and creation operators, respectively, of a harmonic oscillator. Three symbols are widely used nowadays \cite{perelomov86}. First, one can characterize  a hermitean operator $\widehat A$ by its $P$-symbol $P_A ( \alpha )$,  given by the expansion coefficients of $\widehat A$ when expressing it as a linear combination of projectors on coherent states (\ref{CS}):
\be
{\widehat A} = \int_{\sf C} d \mu (\alpha) \, P_A ( \alpha ) \ket{\alpha}\bra{\alpha} \, , \qquad  
        d \mu (\alpha) = \frac{1}{\pi} d\alpha_1 \, d\alpha_2 \, , \quad
        \alpha = \alpha_1 +i \alpha_2 \, .    
\label{psymbol}
\ee
Explicitly, one can write
\be
P_A ( \alpha ) = \sum_{m,n} A_{mn}^{P} \alpha^m                                   \overline{\alpha}^{n} \, 
        \qquad \mbox{ if } \qquad
    {\widehat A} = \sum_{m,n} A_{mn}^{P} a^m (a^+)^{n} \, .
\label{anti}
\ee
Thus, to obtain the P-symbol of ${\widehat A}$, write down the {\em anti-normal} expansion of the operator in terms of annihilation and creation operators and replace them subsequently by $\alpha$ and $\overline{\alpha}$, respectively. Second, the expectation values of $\widehat A$ in the coherent states $\ket{\alpha}$ define its Q-symbol:  
\be
Q_A ( \alpha ) = \bra{\alpha} \, {\widehat A} \, \ket{\alpha} 
         = \sum_{m,n} A_{mn}^{Q} \overline{\alpha}^m \alpha^{n} \, ,
\label{qsymbol}
\ee
which is closely related to the {\em normal-ordered} expansion of the operator. Finally, the Weyl symbol of ${\widehat A}$---providing the Wigner function of a state $\ket{\psi}$ if ${\widehat A} = \ket{\psi}\bra{\psi}$---can be defined in a similar way through {\em symmetrical} ordering \cite{perelomov86}.
\section{Continuous redundant symbols for spin $s$}   
Consider a spin $s$ with Hilbert space ${\cal H}_s$, which carries a $(2s+1)$-dimensional irreducible representation of the group $SU(2)$. The ensemble of observables $\widehat A$ for a spin $s$ is denoted by ${\cal A}_s$, corresponding  to the hermitean $(2s+1)\times (2s+1)$ matrices acting on ${\cal H}_s$. Phase-space symbols of operators are defined in analogy to those for a particle once {\em spin}-coherent states have been introduced. 

Denote the components of the spin operator by $\widehat{\bf S} \equiv \hbar \hat {\bf s}$, satisfying the commutation relations $[{\hat s}_x,{\hat s}_y]= i{\hat s}_z, \ldots$ The standard basis of the space ${\cal H}_s$ is given by the eigenvectors of the $z$ component $ {\hat s}_z = {\bf n}_z \cdot \hat {\bf s} $ of the spin, which are denoted by $\ket{\mu, {\bf n}_z}, -s\leq\mu\leq s$.  The ladder operators $ s_{\pm}= {\hat s}_x {\pm} i {\hat s}_y$ act as usual\footnote{The phases of the states are fixed by the transformation under the anti-unitary time reversal operator $T$:  $T \ket{\mu, {\bf n}_z}=(-1)^{s-\mu} \ket{-\mu,{\bf n}_z}$.} in this basis:
\begin{equation}
{{\hat s}}_ {\pm} \ket{\mu, {\bf n}_z}
            = \sqrt{s(s+1)-\mu(\mu\pm 1)} \, \ket{\mu \pm 1,{\bf n}_z} \, .
\label{splusaction}
\end{equation}
The eigenstates of the operator ${\bf n} \cdot \hat {\bf  s}$ satisfy
\be
{\bf n} \cdot \hat{\bf s} \, \ket{ \mu, {\bf n}} = \mu \, \ket{ \mu, {\bf n}} \, , 
\qquad -s \leq \mu \leq s \, ,
\label{Sneigest}
\ee
where the unit vector ${\bf n}= ( \sin \vartheta \cos \varphi,$ $\sin \vartheta \sin \varphi, \cos \vartheta)$, $0 \leq \vartheta \leq \pi, 0 \leq \varphi < 2\pi$, defines a direction in space. The collection of states with {\em maximal} weight, $\mu = s$, exhausts the {\em coherent} states \cite{arecchi+72},
\be
\ket{{\bf n}} \equiv \ket{s, {\bf n} } 
    = \exp [ -i \, \vartheta \, {\bf m}(\varphi) \cdot \hat{\bf s} \, ] \, 
                       \ket{s,{\bf n}_z} \, ,
\label{defcs}
\ee
with a unit vector ${\bf m}(\varphi) = (- \sin \varphi,\cos\varphi,0)$ in the $xy$ plane. In other words, each coherent state $\ket{ \bf n }$ is obtained from rotating the state $\ket{s,{\bf n}_z}$ about an axis ${\bf m}(\varphi)$ by some angle $\vartheta$. 

For the present purpose, both the Q-symbol and the P-symbol of an operator on ${\cal H}_s$ will be needed; the reader is referred to \cite{varilly+89} for an equivalent of the Weyl symbol, and to \cite{perelomov86,amiet+91} for further details. The P-symbol of an operator $\widehat A$ is introduced in analogy to (\ref{psymbol}) by 
\be
{\widehat A} = \frac{(2s+1)}{4\pi} \int_{{\sf S}^2} d^2{\bf n} P_A ({\bf n}) \ket{\bf n}\bra{\bf n} \, ,   
\label{psymbolspin}
\ee
where the integration is over the surface of the unit sphere, ${\sf S}^2$. An explicit 
expression for the symbol $P_A ({\bf n})$ in terms of the matrix elements $\bra{\mu , {\bf n}_z} {\widehat A} \ket{\mu' , {\bf n}_z}$ can be found in \cite{arecchi+72}. As before, the Q-symbol equals the expectation values of $\widehat A$ in the coherent states (\ref{defcs}),
\be
Q_A ({\bf n}) 
  = \mbox{ Tr } \left[ \, {\widehat A} \, \ket{{\bf n}} \, \bra{{\bf n}} \right]
  = \bra{{\bf n}} \, {\widehat A} \, \ket{{\bf n}}   \, ,
\label{spinQ}
\ee
giving thus rise to a representation of  $\widehat A$ as a function on ${\sf S}^2$, the phase space of a classical spin. The explicit inversion of (\ref{spinQ}) amounts to a straightforward method to reconstruct experimentally a quantum state: if one {\em measures} the expectation values (\ref{spinQ}) of the density matrix $\widehat \rho$ of a spin in all coherent states, the data allow one to determine $\widehat \rho$ unambiguously \cite{manko+97,agarwal98}. 

However, the data (\ref{spinQ}) are highly redundant. As any hermitean operator $\widehat A\in {\cal A}_s$ acting on the space ${\cal H}_s$, the (unnormalized) density matrix $\widehat \rho$ of a spin $s$ depends on $N_s = (2s+1)^2$ real parameters. The symbol $Q_A ({\bf n})$ though takes values on {\em all} points of the unit sphere. Therefore, one is urged to ask whether there are {\em subsets} of the expectation values (\ref{spinQ}) which would permit reconstruction of an operator $\widehat A$ in a more economic way.  This question has been answered in the positive: the determination of $\widehat A$ is possible on the basis of exactly $N_s$ expectation values,  associated with specifically selected directions ${\bf n}_{n}, 1 \leq n \leq N_s$ \cite{amiet+99/2}. In other words, an operator $\widehat A$ is fixed by the values of its Q-symbol at $N_s$ appropriately chosen points; the values of the symbol ``in between'' can be calculated subsequently. For technical reasons, the spatial directions ${\bf n}_{n}$ given in \cite{amiet+99/2} were restricted to a certain class of regular configurations. From now on, a set of $N_s$ points---as well as the 
associated familily of $N_s$ unit vectors ${\bf n}_{n}$---will be referred to as a `constellation' $\cal N$.
\section{Discrete non-redundant symbols for spin $s$}
The purpose of the present paper is to show that the restriction to the specific constellations mentioned above is not necessary: given a generic constellation $\cal M$, the $N_s$ values of the Q-symbol (\ref{spinQ}) contain all the information about the operator $\widehat A$. Let us put it differently: given {\em any} constellation $\cal M$ of vectors ${\bf m}_n$, then {\em either} the numbers $Q_A ({\bf m}_n)$  determine $\widehat A$, {\em or} there is an {\em infinitesimally close} constellation ${\cal M}'$ such that the numbers $Q_A ({\bf m}^\prime_n)$ do the job. Two constellations $\cal M$ and ${\cal M}'$ are close if, for example, the number 
\be 
d({\cal M}' , {\cal M}) = \sum_{n=1}^{N_s} | {\bf m}_{n} - {\bf m}^\prime_{n} | \, ,
\label{distance}
\ee
is small. To visualise this statement, consider the real vector space $I\!\!R^3$: any three unit vectors form a basis provided they are not coplanar or collinear. Among all possibilites, the exceptions have measure zero. At the same time, it is obvious that arbitrarily small variations of the vectors are sufficient to render them linearly independent.   

In the proof given below, the determinant of the Gram matrix of $N_s$ projection operators on coherent states 
\be
{\widehat Q}_n 
= \ket{{\bf m}_{n}} \bra{{\bf m}_{n}}   \, , 
          \qquad  1 \leq n \leq N_s  \, ,
\label{mixedrho}
\ee
is shown to be different from zero for any desired constellation $\cal M$ (or an infinitesimally close one, ${\cal M}'$). Equivalently, one can say that the operators ${\widehat Q}_n$ constitute a {\em quorum} for a spin $s$: they provide a basis for any hermitean operators $\widehat A$:
\begin{equation}
{\widehat A} = \frac{1}{2s+1} \sum_{n=1}^{N_s} A^{n} {\widehat Q}_{n} \, ,
\qquad 
A^{n}  = \mbox{ Tr } \left[ {\widehat A} \, {\widehat {\sf Q}}^{n} \right] \, ;
\label{expandquorum}
\end{equation}
the expansion coefficients $A^{n}$ involve operators ${\widehat {\sf Q}}^{n}$ {\em dual} to the elements of the original basis: $\mbox{ Tr } [ {\widehat Q}_{n} {\widehat {\sf Q}}^{n'}] = \delta_n^{n'}$ \cite{weigert99/1}. 

The $N_s^2$ elements of the Gram matrix ${\sf G}_{nn'}$ \cite{greub63} associated with the constellation $\cal M$ are given by the scalar product of the projectors on coherent states:
\be
 {\sf G}_{nn'} = \mbox{ Tr } \left[ {\widehat Q}_{n} {\widehat Q}_{n'} \right] 
               = | \bra{{\bf m}_{n}} {\bf m}_{n'} \rangle |^2
               = \left( \frac{ 1 + {\bf m}_{n} \cdot {\bf m}_{n'}}{2} \right)^{2s} \, ,
\qquad  1 \leq n,n' \leq N_s \, .
\label{gram}
\ee
It will be essential for the following that the scalar product of two coherent states is a {\em polynomial} in the components of the associated unit vectors ${\bf m}_{n}$ and  ${\bf m}_{n'}$. 

Let us now turn to the proof that $\det {\sf G} \neq 0$ for arbitrary $\cal M$. This can be shown recursively by means of the  matrices ${\sf G} (k), k=1, \ldots, N_s$, which are $(k \times k)$ submatrices of $\sf G$ located in its upper left corner with elements:
\be
{\sf G}_{nn'} (k) = {\sf G}_{nn'} \, , \qquad 1 \leq n,n' \leq k \, .
\label{subG}
\ee
In particular, one has ${\sf G} (1) \equiv 1$, and ${\sf G} (N_s) \equiv {\sf G}$. Suppose that 
the first $(k-1)$ projection operators ${\widehat Q}_n, n=1,\ldots , k-1$, are linearly independent for a specific constellation ${\cal M}_{k-1}$ of $(k-1)$ vectors. It follows that the determinant of ${\sf G} (k-1)$ is different from zero.  Since {\em all} operators on ${\cal H}_s$ do have an expansion of the form (\ref{psymbolspin}), the ensemble of projectors $\{ \ket{\bf n}\bra{\bf n}, {\bf n} \in {\sf S}^2 \} $ spans the {\em entire} space ${\cal A}_s$. Consequently, there is at least one projector ${\widehat Q}_k$ characterized by a vector ${\bf m}^0_k$, say, which cannot be written as a linear combination of the operators ${\widehat Q}_1, \ldots , {\widehat Q}_{k-1}$. Therefore, the determinant of ${\sf G} (k)$ is different from zero at least for this vector ${\bf m}^0_k$. (As it must be, this argument fails if $k$ exceeds $N_s$.) Being a {\em polynomial} function of the $k$-th unit vector, the determinant now will be shown to be different from zero for (almost) any other choice of this vector. It will be possible to select, in particular, the projector associated with the $k$-th vector, ${\bf m}_k$, of the constellation ${\cal M}$---or with a vector ${\bf m}^\prime_k$ infinitesimally close to it: $ | {\bf m}^\prime_k -{\bf m}_k | < \varepsilon/N_s$.   

The determinant of the matrix ${\sf G} (k)$, if conceived as a function of the $k$-th
vector, is infinitely often differentiable with respect to its components, according to (\ref{gram}). Upon keeping the vectors ${\bf m}_1$ to ${\bf m}_{k-1}$ fixed, it may be regarded as a fictitious time-independent {\em Hamiltonian function} of a single classical spin:
\be 
\det {\sf G} ({\bf n}_k) = H_k ({\bf n}_k) \quad (\neq 0 \mbox{ if } {\bf n}_k = {\bf m}^0_k) \, .
\label{hamilton}
\ee
This Hamiltonian describes an {\em integrable} system since there is just one degree of freedom accompanied by one constant of the motion, the Hamiltonian itself cite{arnold84}. The two-dimensional phase space ${\sf S}^2$ is foliated entirely by one-dimensional tori of constant energy. In addition, a finite number of (elliptic or hyperbolic) fixed points and one-dimensional separatrices will occur \. This can be seen, for example, by looking at the flow on the unit sphere generated by $H_k ({\bf n}_k)$:
\be 
\frac{d {\bf n}_k}{dt} = {\bf n}_k \times \frac{\partial H_k}{\partial {\bf n}_k} \, ,
\label{spindyn}
\ee
where $\partial / \partial {\bf n}_k$ is the gradient with respect to ${\bf n}_k$ \cite{srivastava+87}. The right-hand-side is a (non-zero) polynomial in the components of ${\bf n}_k$, implying that the integral curves of the Hamiltonian are fixed points, separatrices, and closed orbits. This means that $H_k ({\bf n}_k)$ can take
the value zero at a finite number of (open or closed) curves or points at most. Consequently, the determinant of ${\sf G} (k)$ is different from zero for almost all choices of ${\bf n}_k$. Therefore, one can rotate smoothly the vector ${\bf m}^0_k$ into any other vector, including ${\bf m}_k$, the $k$-th vector of the desired constellation ${\cal M}$, thereby passing possibly through points with $\det {\sf G} (k) = 0$. If, accidentally,  ${\bf m}_k$ corresponds to a point with vanishing energy (this happens with probability zero only), one can nevertheless approach it arbitrarily close by a vector ${\bf m}^\prime_k$ with $| {\bf m}^\prime_k - {\bf m}_k | < \varepsilon /N_s$ since levels of constant energy have a co-dimension at most equal to one. 

Working one's way from $k=2$ to $N_s$, one ends up with a constellation ${\cal M}'$ which is infinitesimally close to ${\cal M}$ since $\sum_n | {\bf m}^\prime_n - {\bf m}_n | < \varepsilon$ can be made arbitrarily small; with probability one, however, the constellation $\cal M$ is obtained exactly. Consequently, almost all constellations ${\cal M}$ of $N_s$ projection operators ${\widehat Q}_{n}$ give rise to a basis in the space of linear operators on ${\cal H}_s$. In turn, the values of the {\em discrete} Q-symbol related to a constellation $\cal M$ are indeed sufficient to determine the operator $\widehat A$.  
\section{Discussion and Outlook}
It has been shown that (almost) any distribution of $N_s$ points on the sphere ${\sf S}^2$ gives rise to a non-orthogonal basis of coherent-state projectors ${\widehat Q}_n$ in the linear space ${\cal A}_s$ of operators for a spin $s$. The collection of expectation values of an operator $\widehat A$ in these states will be called its {\em discrete} phase-space symbol, containing no redundant information. The expansion (\ref{expandquorum}) is the discrete analogue of (\ref{psymbolspin}) so that the coefficients $A^n$ correspond to the {\em discrete} P-symbol of $\widehat A$, which also represents it in a non-redundant way. A second expansion in the basis $\widehat{\sf Q}^n$ dual to the basis ${\widehat Q}_n$
requires the discrete Q-symbol as coefficients:
\begin{equation}
{\widehat A} = \frac{1}{2s+1} \sum_{n=1}^{N_s} Q_A ({\bf n}_n) {\widehat {\sf Q}}^{n} \, ,
\qquad 
Q_A ({\bf n}_n)  = \mbox{ Tr } \left[ {\widehat A} \, {\widehat Q}_{n} \right] \, .
\label{expandquorumdual}
\end{equation}
Therefore, discrete Q- and P-symbols can be considered as dual to each other, providing co- and contravariant coordinates of the operator $\widehat A$. Contrary to the redundant P-symbol (\ref{psymbolspin}), the discrete one is unique \cite{gilmore76}. 

For a particle system, it is natural to ask a closely related question: which subsets of coherent-state projectors form a basis in the space of {\em operators} acting on the particle Hilbert-space? Minimal bases for the Hilbert space, i.e. pure particle {\em states}, have been identified by von Neumann \cite{perelomov86}. A matrix representation of $\widehat A$ with respect to an orthogonal basis of the Hilbert space depends on a countable number of parameters while the Q-symbol $Q_A$ is a smooth function in the complex plane. Therefore, one might conjecture that the values $Q_A $ on some countable subset of the complex plane should suffice to determine $\widehat A$. The expansion coefficients of $\widehat A$ in such a non-redundant basis would 
then correspond to its non-redundant P-symbol. 
\subsection*{Acknowledgements}
St. W. acknowledges financial support by the {\em Schweizerische Nationalfonds}.  
\end{document}